\documentclass[aps,article]{revtex4}

\usepackage{epsfig}

\begin{document}
\title{{\bf Correlations in interfering electrons irradiated by nonclassical microwaves}}

\author{C.C. Chong, D.I. Tsomokos, A. Vourdas}

\affiliation{Department of Computing,
School of Informatics,\\
University of Bradford,\\
Bradford BD7 1DP, United Kingdom}

\begin{abstract}
Electron interference in mesoscopic devices irradiated by external
non-classical microwaves, is considered. In the case of one-mode microwaves,
it is shown that both the average intensity and the spectral density of the
interfering electrons are sensitive to the quantum noise of the microwaves.
The results for various quantum states of the microwaves are compared and
contrasted with the classical case. Separable and entangled two-mode
microwaves are also considered and their effect on electron average intensity
and autocorrelation, is discussed.
\end {abstract}

\emph{Phys. Rev.} \textbf{A66}, 033813 (2002).


\maketitle

\section{Introduction}
Interference of electrons that encircle a magnetostatic flux has been studied
for a long time since the pioneering work of Aharonov and Bohm [1]. It has
applications in various contexts, for example in conductance oscillations in
mesoscopic rings [2], neutron interferometry [3], and `which-path' experiments
[4].

A further development is the replacement of the magnetostatic flux with an
electromagnetic field. In this case, the electrons feel not only the vector
potential but the electromagnetic field. Therefore, the objective in this `ac
Aharonov-Bohm' effect is very different from the `dc Aharonov-Bohm' effect
(with magnetostatic flux). In the latter case the physical reality of the
vector potential has been demonstrated and the subtleties of quantum mechanics
in non-trivial topologies have been studied. The former case constitutes a
non-linear device, where the interaction between the interfering electrons and
the photons leads to interesting non-linear phenomena [5] like amplification
and frequency conversion. Indeed the non-linearity can be seen in the
intensity of the interfering electrons which is a sinusoidal function of the
time-dependent magnetic flux. Our study is related to recent work on the
interaction of mesoscopic devices with microwaves [6].

One step further in this line of work is to use in these experiments
non-classical microwaves, where the quantum noise is carefully controlled. In
this case, we can quantify how the quantum noise destroys slightly the
electron interference [7]. More generally, we have here two coupled quantum
systems (photons and electrons) and we can study how various quantum phenomena
associated with the non-classical electromagnetic field cause corresponding
quantum phenomena on the electrons. For example, we can study how the quantum
statistics of photons affects the quantum statistics of the electrons; how the
entanglement of two-mode non-classical microwaves affects the electron
interference; etc.

In this paper we study how various types of non-classical microwaves affect
both the average intensity and the spectral density of the interfering
electrons. The results are compared with those for classical microwaves
(Section II). We also consider two-mode microwaves with frequencies $\omega_1$
and $\omega_2$ in Section III, where we show that we get different results for
rational and irrational values of $\omega_1/\omega_2$. We interpret these
results in terms of emission and absorption of photons by the non-linear
device of the interfering electrons. This discussion shows the potential use
of the device for frequency conversion. Two-mode microwaves can be
factorizable, separable or entangled [8]. We study how such deep quantum
phenomena in the microwaves can affect the electron interference. The problem
is complex and it is approached through examples which demonstrate the effect.
In particular, we compare and contrast the effect on electron interference, of
an entangled microwave state with that of the corresponding separable
microwave state. We conclude in Section IV with a discussion of our results.

\section{One-mode microwaves}
\subsection{Classical microwaves}
Interfering electric charges in mesoscopic devices that follow two different
paths $C_0$ and $C_1$ are considered. A magnetic flux $\phi$ is threading the
surface between the two paths. This is referred to as the dc or ac
Aharonov-Bohm experiment, according to whether the magnetic flux is
time-independent or time-dependent, correspondingly. In the dc Aharonov-Bohm
experiment the electric charges feel only a vector potential. In the ac
Aharonov-Bohm experiment the electric charges also feel an electric field,
which is induced according to Faraday's law. The ac Aharonov-Bohm effect can
be realised experimentally: at low frequencies using a solenoid with a
suitable time-dependent current; or at high frequencies using a waveguide,
whose magnetic and electric fields are perpendicular and parallel to the plane
of the two paths, respectively (Fig. 1).

Let $\psi _0$, $\psi _1$ be the electron wavefunctions with winding numbers
$0, 1$ correspondingly, in the absence of magnetic field. The effect of the
electromagnetic field is the phase factor $\exp [ie\phi (t)]$ and the
intensity is
\begin{equation}
I(t)=|\psi _0+\psi _1\exp [ie\phi (t)]|^2=|\psi _0|^2+|\psi _1|^2
+2|\psi _0||\psi _1|\Re \{\exp [i(\sigma+e\phi(t))]\}
\end{equation}
where $\sigma = \mbox{arg}(\psi _1)-\mbox{arg}(\psi _0)$. Units in which
$k_B=\hbar=c=1$ are used throughout. For simplicity we consider the case of
equal splitting, in which $|\psi _0|^2=|\psi _1|^2=1/2$ and let $\sigma =0$.
In this case we get
\begin{equation} \label{I_cl}
I(t) = 1 + \cos[e \phi(t)].
\end{equation}

\begin{figure} [h]
\centering \scalebox{0.6} {\includegraphics{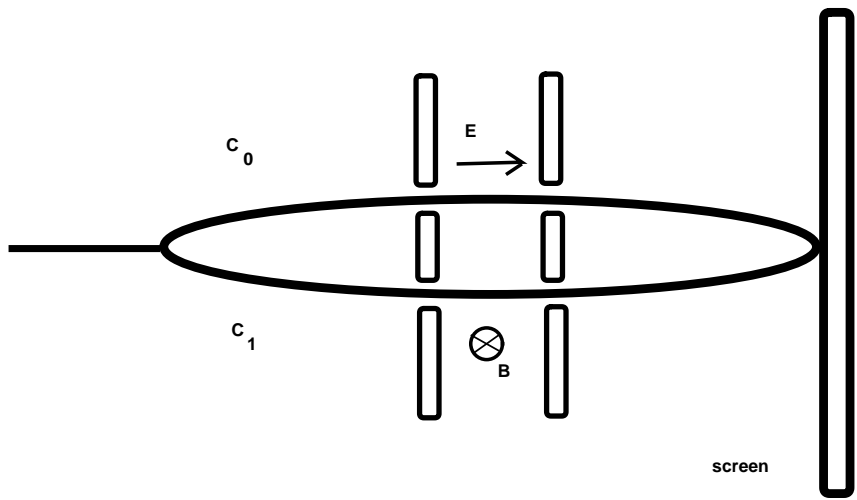}}\caption{ac
Aharonov-Bohm phenomenon where electrons interfere in the presence of
time-dependent magnetic flux (electromagnetic field). The electromagnetic
field travels in the waveguide shown, with the electric field parallel to the
plane of the diagram and the magnetic field perpendicular to it. The electrons
follow the paths $C_0$, $C_1$ as shown.}
\end{figure}

In general, for a complex intensity $I(t)$ the autocorrelation function is
defined as
\begin{equation}\label{autocorr}
\Gamma(\tau) = \lim_{T\rightarrow \infty} \frac{1}{2T}
\int_{-T}^{T} R(t,\tau) dt;\;\;\;\;\;\;R(t,\tau)\equiv I^{*}(t)I(t+\tau).
\end{equation}
The following properties of the autocorrelation function are well known:
\begin{eqnarray} \label{pro1}
\Gamma(-\tau)=\Gamma^{*}(\tau);\;\;\;\;\;
\Gamma(0)\geq 0 ;\;\;\;\; \;
|\Gamma(\tau)|\leq \Gamma(0) .
\end{eqnarray}
It will be explained later that these relations are also true in the case of
non-classical microwaves. The normalized autocorrelation function is defined
as
\begin{eqnarray} \label{normalG}
\gamma(\tau)=\frac{\Gamma(\tau)}{\Gamma(0)};\;\;\;\;\;
0\leq|\gamma(\tau)|\leq1.
\end{eqnarray}
For one-mode microwaves, the autocorrelation function $\Gamma(\tau)$ for the
charges will be periodic with a period $2\pi/\Omega$, where $\Omega$ is, by
definition, the frequency associated with the periodic function
$\Gamma(\tau)$. An expansion of $\Gamma(\tau)$ into a Fourier series gives the
spectral density $S_K$:
\begin{eqnarray} \label{psd}
S_K &=& \frac{\Omega}{2\pi}\int_{0}^{2\pi/\Omega}\Gamma(\tau)\exp(-iK\Omega\tau) d\tau \nonumber \\
\Gamma(\tau) &=&\sum_{K=-\infty}^{\infty}S_K\exp(iK\Omega\tau).
\end{eqnarray}
Equation (\ref{pro1}) implies that the coefficients $S_K$ are real numbers,
for both classical and non-classical microwaves.

We consider the case where the classical time-dependent flux is given by
\begin{equation}
\phi(t) = \phi_1\sin(\omega t)
\end{equation}
and using Eqs.(\ref{I_cl}),(\ref{autocorr}) we find
the autocorrelation function:
\begin{equation} \label{G_cl}
\Gamma_{cl}(\tau)=\left[1 + J_{0}(e \phi_1)\right]^2 +
2\sum_{K=1}^{\infty}\left[J_{2K}(e \phi_1) \right]^2 \cos(2K\omega\tau),
\end{equation}
where $J_K$ are Bessel functions.
Comparison of Eqs.(\ref{psd}),(\ref{G_cl}) shows that $\Omega =2\omega$ and
\begin{equation}
S_0=[1 + J_{0}(e \phi_1)]^2;\;\;\;\;\;\;S_K=[J_{2K}(e \phi_1)]^2.
\end{equation}
We note that $S_K=S_{-K}$. This is because in the classical case considered in
this section, $I(t)$ is real and consequently $\Gamma(\tau)$ is real.
Therefore Eq.(\ref{pro1}) shows that $\Gamma(\tau)$ is an even function, which
implies that $S_K=S_{-K}$. It is stressed that in the non-classical case
considered next, $\Gamma(\tau)$ is complex in general and $S_K\ne S_{-K}$.

\subsection{Non-classical microwaves}

A monochromatic electromagnetic field of frequency $\omega$ is considered, at
temperatures $k_B T<<\hbar \omega$. In quantized electromagnetic fields the
vector potential $A_i$ and the electric field $E_i$ are dual quantum
variables. For a loop $C=C_0-C_1$ (where $C_0$ and $C_1$ are the paths
corresponding to 0,1 winding), which is small in comparison to the wavelength
of the microwaves, the $A_i$ and the $E_i$ can be integrated around $C$ and
yield the magnetic flux $\phi =\oint _C A_i dx_i$ and the electromotive force
$V_{\rm EMF}=\oint _C E_i dx_i$, correspondingly, as dual quantum variables.
The size of a mesoscopic device is usually of the order of $0.1 \mu m$ and is
indeed much smaller than the microwave wavelength. The annihilation operator
is now introduced as $a=2^{-\frac{1}{2}}\xi^{-1}\left( \phi+i\omega^{-1}V_{\rm
EMF}\right)$ and the corresponding creation operator, where $\xi$ is a
constant proportional to the area enclosed by $C$. The flux operator is
consequently written as $\phi(t)= \exp (itH)\phi(0)\exp (-itH)$, where $H$ is
the Hamiltonian that contains the $\omega a^\dagger a$ term and an interaction
term. In the `external field approximation' the interaction term, which
describes the back-reaction from the electrons on the electromagnetic field,
is neglected. This is a good approximation for external fields which are
strong in comparison to those produced dynamically by the currents in the
mesoscopic device (back-reaction). In this approximation the interaction term
can be ignored and we get
\begin{equation}
\hat \phi(t)=\frac{\xi}{\sqrt{2}} \left[\exp(i\omega t)a^\dagger + \exp(-i\omega t)a\right].
\end{equation}

The phase factor $\exp(ie\phi)$ is now the operator
\begin{equation}
\exp\left[ie\hat \phi(t)\right]=D\left[iq\exp(i\omega t)\right], \ \ \ q=\frac{\xi e}{\sqrt{2}}
\end{equation}
where $D(\lambda)$ is the displacement operator
$D(\lambda) = \exp (\lambda a^\dagger - \lambda ^*a)$.
The interference between the two electron beams is described by the intensity operator
\begin{eqnarray}
\hat{I}(t) = 1 + \cos\left[e \hat{\phi}(t)\right] = 1 +
\frac{1}{2} D\left[iq\exp(i\omega t)\right ]
+\frac{1}{2} D\left[-iq\exp(i\omega t)\right ].
\end{eqnarray}
Let $\rho$ be the density matrix describing the external non-classical microwaves.
We can now calculate the expectation value of the electron intensity
\begin{equation} \label{I_ave}
\langle I(t) \rangle \equiv \mbox{Tr}\left[\rho\hat{I}(t)\right]=
1+\frac{1}{2}\tilde W(\lambda )+\frac{1}{2}\tilde W(-\lambda ); \ \ \ \
\lambda=iq\exp(i\omega t),
\end{equation}
where $\mbox{Tr}\left[\rho D(\lambda)\right]\equiv \tilde W(\lambda)$ is the
Weyl (or characteristic) function which has been studied by various authors
including ourselves (e.g. [9] and references therein). The tilde in the
notation reflects the fact that the Weyl function is related to the Wigner
function through a two-dimensional Fourier transform. Physically the ${\rm
Tr}\left[\rho \hat I(t)\right]$ describes the exchange of photons between the
electrons and the external electromagnetic field. Expansion of the
exponentials in Eq.(13) gives an infinite sum of terms of the type ${\rm
Tr}\left[\rho (ae^{-i\omega t})^N (a^{\dagger}e^{i\omega t})^M \right]$ which
describe processes in which the electrons emit $M$ photons to the external
electromagnetic field and at the same time absorb $N$ photons from the
external electromagnetic field. Summation of the appropriate coefficients
leads to Bessel functions which appear in most of the calculations throughout
the paper. We note that a similar expansion and a similar interpretation can
also be made in the classical microwave case. However, in this case instead of
creation and annihilation operators we have classical numbers and the
interpretation is perhaps less convincing.

\begin{figure} [h]
\centering \scalebox{0.6} {\includegraphics{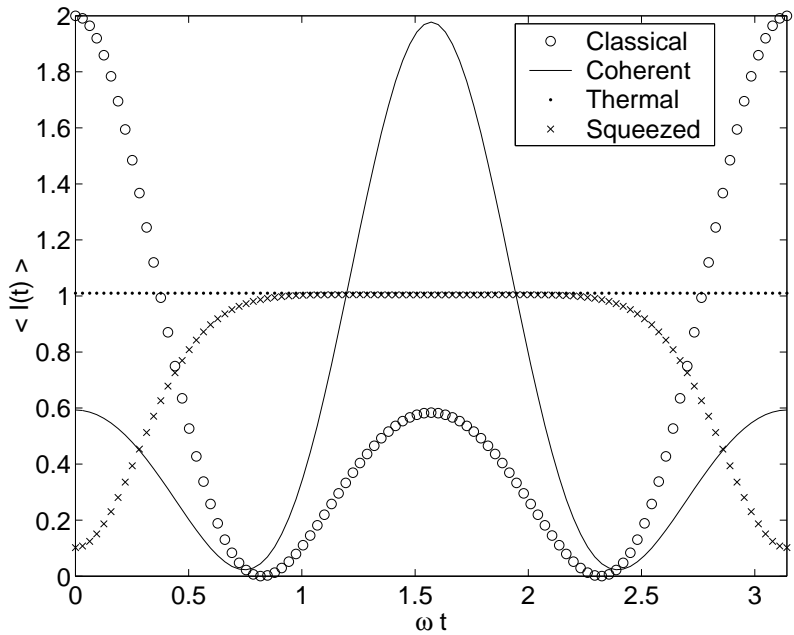}}\caption{$\langle
I(t)\rangle$ as a function of $\omega t$ for $\omega=10^{-4}$, $\langle N
\rangle =100$, $r=5.5$. We use units where $\hbar=k_B=c=1$.}
\end{figure}

Ref.[10] has considered several density matrices and presented results for $\tilde W(\lambda)$.
Using them we have calculated
$\langle I(t) \rangle$ for various quantum states of the microwaves.
We are also interested in the quantity
\begin{equation} \label{R_exp}
R(t,\tau)\equiv \mbox{Tr}\left[\rho\hat{I}^{\dagger}(t)\hat{I}(t+\tau)\right]
\end{equation}
which is calculated for various density matrices, as well as $\Gamma(\tau)$
using Eq.(\ref{autocorr}). The Fourier series of Eq.(\ref{psd}) leads to the
coefficients $S_K$.

We note that using the relation
\begin{equation} \label{Gamma_tau}
\Gamma(\tau) = \lim_{T\rightarrow \infty} \frac{1}{2T}
\int_{-T}^{T} \mbox{Tr}\left[\rho\hat{I}^{\dagger}(t)\hat{I}(t+\tau)\right] dt
\end{equation}
in conjunction with the fact that for any operator $\hat{O}$,
$\mbox{Tr}\left(\hat{O}^\dagger\right)=\left[\mbox{Tr}(\hat{O}) \right]^{*}$,
we prove $\Gamma(-\tau)=\Gamma^{*}(\tau)$ and therefore the coefficients $S_K$
are real numbers. As we already pointed out, $\Gamma(\tau)$ is in general
complex. This is intimately related with the fact that the operators
$\hat{I}^{\dagger}(t)$ and $\hat{I}(t+\tau)$ do not commute. In fact, the
imaginary part of $R(t,\tau)$ is
$(1/2)\mbox{Tr}\{\rho[\hat{I}(t),\hat{I}(t,\tau)]\}$. In the classical case,
these quantities are not operators, they commute and consequently $R(t,\tau)$
is real.

\subsubsection{Microwaves in coherent states}

For coherent states
$|A\rangle =D(A)|0\rangle$ the $R(t,\tau)$ is
\begin{eqnarray}
R_{coh}(t,\tau)&=& 1 + \exp\left(-\frac{q^2}{2}\right)\cos\left[2q |A|\cos(\omega t - \theta_A)\right]\nonumber\\
& &+\exp\left(-\frac{q^2}{2}\right)\cos\left[2q |A|\cos(\omega t + \omega\tau - \theta_A)\right]\nonumber\\
& &+\frac{1}{2}\exp\left\{-q^2[1+\exp(i\omega\tau)]\right\}\cos\left[4q\cos\left(\frac{\omega\tau}{2}\right)|A|\cos\left(\omega t + \frac{\omega\tau}{2}-\theta_A\right)\right] \nonumber\\
& &+ \frac{1}{2}\exp\left\{-q^2[1-\exp(i\omega\tau)]\right\}\cos\left[4q\sin\left(\frac{\omega\tau}{2}\right)|A|\sin\left(\omega t + \frac{\omega\tau}{2}-\theta_A\right)\right]
\end{eqnarray}
where $\theta_A =\mbox{arg} (A)$. Using Eq.(15) the electron autocorrelation
function $\Gamma(\tau)$ for microwaves in coherent states is found as
\begin{eqnarray}
\Gamma_{coh}(\tau)
&=&1 + 2\exp\left(-\frac{q^2}{2}\right)J_0\left(2q|A|\right)\nonumber\\
&&+\frac{1}{2}\exp\left\{-q^2\left[1-\exp(i\omega\tau)\right]\right\}
J_0\left[4q\sin\left(\frac{\omega\tau}{2}\right)|A|\right]\nonumber\\
&&+\frac{1}{2}\exp\left\{-q^2\left[1+\exp(i\omega\tau)\right]\right\}
J_0\left[4q\cos\left(\frac{\omega\tau}{2}\right)|A|\right].
\end{eqnarray}
In contrast to the case of classical microwaves, $\Gamma_{coh}(\tau)$ is now a
complex function and $S_K \ne S_{-K}$. This is a periodic function with period
$\pi/\omega$ and a Fourier series analysis is performed numerically as in
Eq.(\ref{psd}).

\subsubsection{Microwaves in squeezed states}
Squeezed states are defined as
\begin{eqnarray}
|B;r\vartheta\rangle &=& S(r\vartheta)|B\rangle =S(r\vartheta )D(B)|0\rangle \\
S(r\vartheta)&=&\exp\left[-\frac{r}{4}\exp(-i\vartheta)a^{\dagger 2}+ \frac{r}{4}\exp(i\vartheta)a^2 \right].
\end{eqnarray}
where $S(r\vartheta)$ is the squeezing operator. The expectation value for the
electron $R_{sq}(t,\tau)$ is given by
\begin{eqnarray} \label{G_sq}
R_{sq}(t,\tau)&=&1 + \exp(-Y_1)\cos(X_1) + \exp(-Y_2)\cos(X_2) \nonumber\\
& &+\frac{1}{2}\exp\left[-iq^2\sin(\omega \tau)\right]\exp(-Y_3)\cos(X_3)\nonumber\\
&& + \frac{1}{2}\exp\left[iq^2\sin(\omega \tau)\right]\exp(-Y_4)\cos(X_4)
\end{eqnarray}
where the $Y_j$ and $X_j$, are given in the Appendix. Using this result we
have calculated the $\Gamma_{sq}(\tau)$ numerically. It can easily be verified
that for $r=0$ the squeezed states results reduce to the coherent states
results. $\Gamma_{sq}(\tau)$ is a periodic function with period $\pi/\omega$
and a Fourier series analysis is performed numerically as in Eq.(\ref{psd}).

\subsubsection{Microwaves in thermal states}

For thermal states, the $R(t,\tau)$ is
\begin{eqnarray}
R_{th}(t,\tau)&=&1 +
2\exp\left[-\frac{q^2}{2}\coth\left(\frac{\beta\omega}{2}\right)\right] \nonumber\\
&&+\frac{1}{2}\exp\left[iq^2\sin(\omega\tau)- 2q^2\sin^2\left(\frac{\omega\tau}{2}\right)
\coth\left(\frac{\beta\omega}{2}\right)\right] \nonumber \\
&&+\frac{1}{2}\exp\left[-iq^2\sin(\omega\tau)-2q^2\cos^2\left(\frac{\omega\tau}{2}\right)
\coth\left(\frac{\beta\omega}{2}\right)\right]
\end{eqnarray}
and clearly $\Gamma_{th}(\tau)=R_{th}(\tau)$. The $\Gamma_{th}(\tau)$ is a
periodic function with period $\pi/\omega$ and its Fourier coefficients are
calculated numerically.

\begin{figure} [h]
\centering \scalebox{0.6}
{\includegraphics{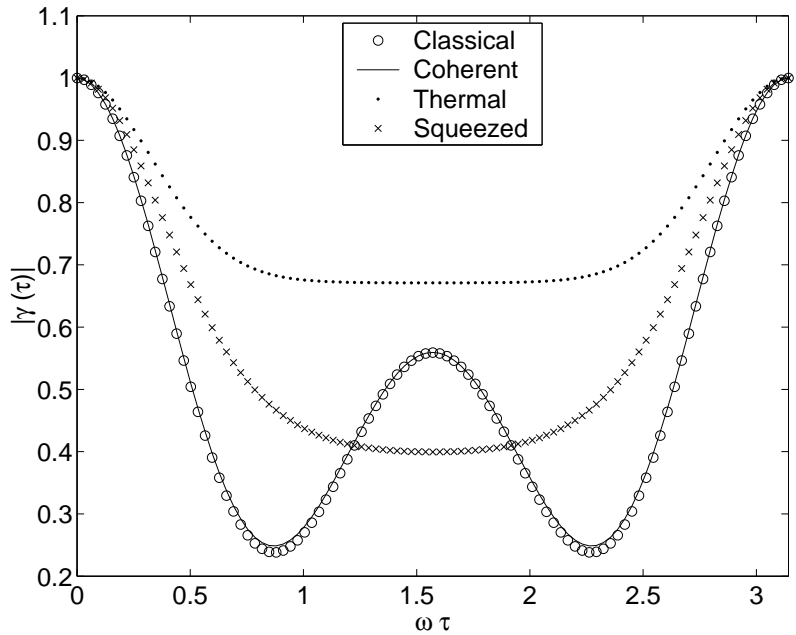}}\caption{$|\gamma(\tau)|$ as a function of $\omega
\tau$ for $\omega=10^{-4}$, $\langle N \rangle =100$, $r=5.5$. We use units
where $\hbar=k_B=c=1$.}
\end{figure}

\subsection{Results}
Numerical results are presented for the four cases:
classical microwaves and non-classical microwaves in coherent, squeezed and thermal states.
For a meaningful comparison, we consider the case where
the average number of photons $\langle N \rangle$ in coherent, squeezed and thermal states
is the same:
\begin{eqnarray}
\langle N \rangle =|A|^2&=& \left[\sinh \left(\frac{r}{2}\right)\right]^2 +
\left[\cosh \left(\frac{r}{2}\right)
- \sinh\left(\frac{r}{2}\right) \right]^2 B^2 \nonumber\\
&=&\frac{1}{\exp(\beta \omega) -1}.
\end{eqnarray}
For the classical case we took $\phi_1^2 = 2|A|^2=2\langle N\rangle$. In all
results of Figs. 2 to 5, $\omega=10^{-4}$ (which in our units is $eV$),
$\langle N \rangle =100$, $r=5.5$.

Fig. 2 shows the $\langle I(t)\rangle$ as a function of $\omega t$. In Fig. 3,
the absolute value of the normalized autocorrelation function $|\gamma(\tau)|$
[Eq.(\ref{normalG})] is shown as a function of $\omega \tau$. The period of
$|\gamma(\tau)|$ is $\pi/\omega$ (i.e., $\Omega=2\omega$) and the plots are
presented from $0$ to $\pi$. As explained earlier the $\gamma(\tau)$ is real
in the case of classical microwaves, but it is complex in general in the case
of non-classical microwaves. This is shown explicitly in Fig. 4, which
includes the imaginary parts of $\gamma(\tau)$ for all cases, as a function of
$\omega \tau$. Fig. 5 shows the Fourier coefficients $S_K$ ($K=-2,...,2$).

\begin{figure} [h]
\centering \scalebox{0.6} {\includegraphics{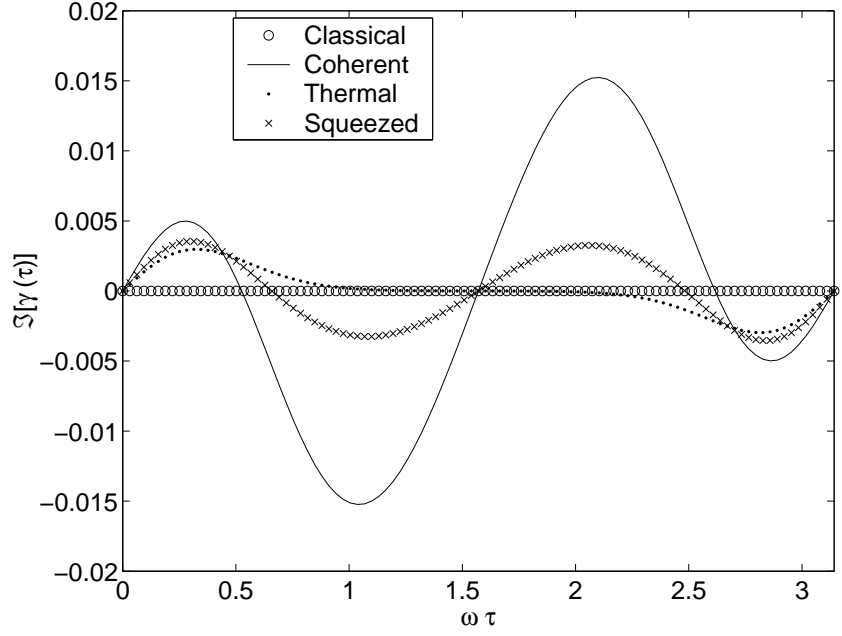}}\caption{$\Im
[\gamma(\tau)]$ as a function of $\omega \tau$ for $\omega=10^{-4}$, $\langle
N \rangle =100$, $r=5.5$. We use units where $\hbar=k_B=c=1$.}
\end{figure}

The results quantify the effect of quantum noise on interference. All
microwaves that we have considered have the {\bf same average number of
photons} and they differ in the quantum noise. For the classical microwaves
(where the concept of number of photons is not applicable) the amplitude is
equal to the amplitude of the microwaves in the coherent state. These four
types of microwaves lead to different electron interference results. Fig. 2
shows clearly that $\langle I(t)\rangle$ is different in all these cases. Fig.
3 shows that the absolute normalized electron autocorrelations are different,
with the exception of the classical result which is almost identical to the
coherent result. The imaginary part of the electron autocorrelation (Fig. 4)
distinguishes the classical from the non-classical microwave cases. It is zero
for classical microwaves and takes various distinct non-zero values for
different types of non-classical microwaves. The same effect can also be seen
through the spectral density coefficients $S_K$ in Fig. 5 which are simply the
Fourier transform of the electron autocorrelation function (Eq.(6)).

\begin{figure} [h]
\centering \scalebox{0.6} {\includegraphics{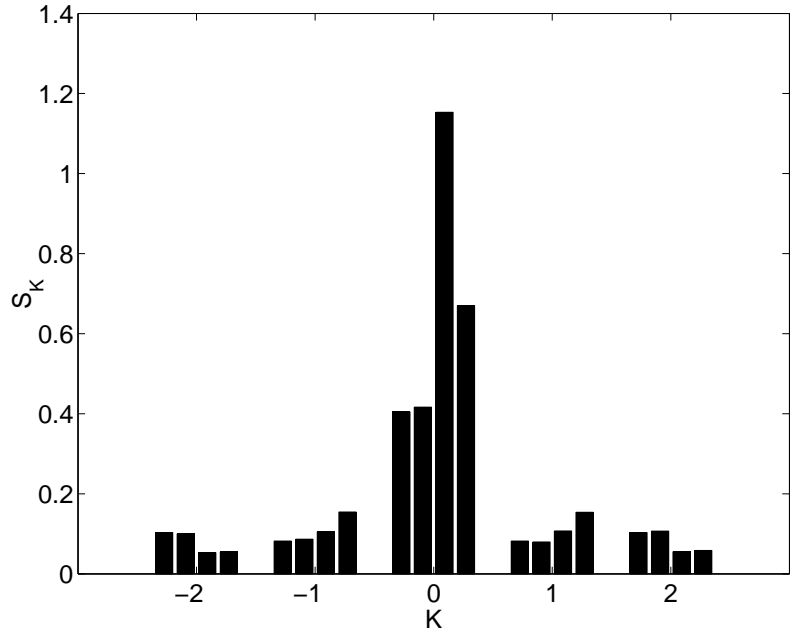}}\caption{$S_K$
($K=-2,...,2$) for $\omega=10^{-4}$, $\langle N \rangle =100$, $r=5.5$. The
four columns of each value of $K$ represent from left to right classical,
coherent, thermal and squeezed microwaves. We use units where
$\hbar=k_B=c=1$.}
\end{figure}

\newpage
\section{Two-mode microwaves}

\subsection{Classical microwaves}

The case of classical two-mode microwaves
\begin{eqnarray}
\phi(t) = \phi_1\sin(\omega _1 t)+\phi_2\sin(\omega _2t).
\end{eqnarray}
is considered. In this case Eq.(\ref{I_cl}) gives the electron intensity
\begin{equation} \label{I_2mode}
I(t) = 1 + \cos[e\phi_1\sin(\omega _1 t)+e\phi_2\sin(\omega _2t) ],
\end{equation}
which is a periodic function. The autocorrelation function is different in the
two cases where the ratio $\omega_1 / \omega_2$ takes rational and irrational
values. The physical reason for this is that in the rational case, where
$\omega_1/\omega_2=P/Q$ and $P, Q$ are coprime integers, the non-linear system
can act as a frequency converter by absorbing $Q$ photons of frequency
$\omega_1$ and emitting $P$ photons of frequency $\omega_2$. The relation
$Q\omega_1 = P\omega_2$ expresses the conservation of energy. In the
irrational case, the system cannot act as a frequency converter simply because
there is no analogous relation for the conservation of energy.

Combining Eqs.(\ref{autocorr}),(\ref{I_2mode}) it is found that in the case of
irrational $\omega_1/\omega_2$, the autocorrelation is
\begin{equation} \label{G_ira}
\Gamma_{ir}(\tau) = 1 +2J_0(e\phi_1)J_0(e\phi_2)
+\sum_{n,k=-\infty}^{\infty} \mu(\tau) \left[J_n(e\phi_1)\right]^2\left[J_{2k-n}(e\phi_2)\right]^2
\end{equation}
where $\mu(\tau) = \exp\{-i[n\omega_1 + (2k-n)\omega_2]\tau\}$.
In the case that the ratio $\omega_1/\omega_2=P/Q$ (rational), the autocorrelation is
\begin{eqnarray} \label{G_ra}
\Gamma_{ra}(\tau) &=& 1+
\sum_{n=-\infty}^{\infty}J_{Qn}(e\phi_1)J_{-Pn}(e\phi_2)
+ \sum_{n=-\infty}^{\infty}J_{Qn}(-e\phi_1)J_{Pn}(e\phi_2)\nonumber\\
& &+\frac{1}{4}\sum_{n,m,N = -\infty}^{\infty}\nu(\tau) J_n (e\phi_1)J_m (e\phi_2)J_{NQ-n}(e\phi_1)J_{NP-m}(e\phi_2)\nonumber\\
& &+\frac{1}{4}\sum_{n,m,N = -\infty}^{\infty}\nu(\tau) J_n (e\phi_1)J_m (e\phi_2)J_{NQ-n}(-e\phi_1)J_{NP-m}(-e\phi_2)\nonumber\\
& &+\frac{1}{4}\sum_{n,m,N = -\infty}^{\infty}\nu(\tau) J_n (-e\phi_1)J_m (-e\phi_2)J_{NQ-n}(e\phi_1)J_{NP-m}(e\phi_2)\nonumber\\
& &+\frac{1}{4}\sum_{n,m,N = -\infty}^{\infty}\nu(\tau) J_n(-e\phi_1)J_m (-e\phi_2)J_{NQ-n}(-e\phi_1)J_{NP-m}(-e\phi_2)
\end{eqnarray}
where $\nu(\tau) = \exp[i(NQ-n)\omega_1\tau +i(NP-m)\omega_2\tau]$. It is
interesting to explain the results of Eqs.(\ref{G_ira}),(\ref{G_ra}) taking
into account the interpretation of the expansion of the exponentials in terms
of emission/absorption of photons (discussed after Eq. (13)) in conjunction
with the above comments about frequency conversion. For example, in the last
term of Eq.(\ref{G_ra}) for the rational case, the system emits $n$ photons of
frequency $\omega_1$ at time $t$ (related to an exponential $\exp(in\omega_1
t)$); emits $NQ-n$ photons of frequency $\omega_1$ at time $(t+\tau)$ (related
to an exponential $\exp[i(NQ-n)\omega_1 (t+\tau)]$); absorbs $m$ photons of
frequency $\omega_2$ at time $t$ (related to an exponential $\exp(-im\omega_2
t)$); and absorbs $NP-m$ photons of frequency $\omega_2$ at time $(t+\tau)$
(related to an exponential $\exp[-i(NP-m)\omega_2 (t+\tau)]$). Taking into
account the relation $\omega_1/\omega_2=P/Q$ we see that the product of these
exponentials is the factor $\nu(\tau)$. Similarly, in the last term of
Eq.(\ref{G_ira}) for the irrational case the system emits $n$ photons of
frequency $\omega_1$ at time $t$; absorbs $n$ photons of frequency $\omega_1$
at time $(t+\tau)$; absorbs $(2k-n)$ photons of frequency $\omega_2$ at time
$t$; and emits $(2k-n)$ photons of frequency $\omega_2$ at time $(t+\tau)$. In
this case there is no transfer of energy (frequency conversion) between the
two frequencies. As previously, the factor $\mu(\tau)$ is related to the
exponentials associated with the absorption/emission of photons. Clearly, the
electron autocorrelation is a periodic function of $\tau$ only in the rational
case.

\subsection{Entangled two-mode microwaves}
We next consider non-classical two-mode microwaves. We are particularly
interested to study how entangled two-mode microwaves affect the electron
interference. For this reason we consider the entangled state $|s\rangle
=2^{-1/2}(|01\rangle +|10\rangle )$ where $|01\rangle$ , $|10\rangle$ are two
mode number eigenstates. For comparison we also consider the separable
(disentangled) state
\begin{equation} \label{sep}
\rho _{sep}=\frac {1}{2} (|01\rangle \langle 01| +|10\rangle \langle 10|).
\end{equation}
Clearly, the density matrix of the entangled state $\rho _{ent}=|s\rangle \langle s|$ can be written as
\begin{equation} \label{ent}
\rho _{ent}=\rho _{sep}+\frac {1}{2} (|01\rangle \langle 10| +|10\rangle \langle 01|).
\end{equation}
In this case using Eq.(\ref{I_ave}) with
\begin{eqnarray}
\hat{I}(t) = 1 + \frac{1}{2} D_1(\lambda_1)D_2(\lambda_2)+ \frac{1}{2} D_1(-\lambda_1)D_2(-\lambda_2); \ \ \ \
\lambda_j = iq\exp(i\omega_j t)
\end{eqnarray}
for two modes ($j=1,2$) we find that
\begin{eqnarray}
\langle I(t)\rangle_{sep}&=& 1 + \left(1-q^2\right)\exp\left(-q^2\right), \\
\langle I(t)\rangle_{ent}&=& \langle I(t)\rangle_{sep} - q^2\exp\left(-q^2\right)\cos[(\omega_1 -\omega_2)t].
\end{eqnarray}

\begin{figure} [h]
\centering \scalebox{0.6} {\includegraphics{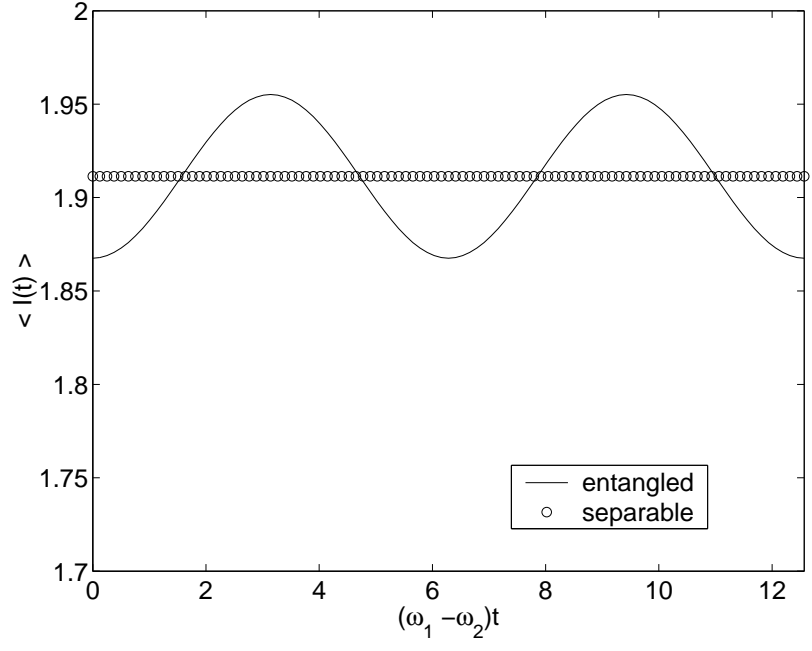}}\caption{$\langle
I(t)\rangle$ as a function of $t(\omega_1 - \omega_2)$ for the separable and
entangled cases of Eqs.(\ref{sep}) and (\ref{ent}). We use units where
$\hbar=k_B=c=1$.}
\end{figure}

These results are presented in Fig. 6. It is seen that for the example we
considered, the $\langle I(t)\rangle _{sep}$ is constant in time, while the
$\langle I(t)\rangle_{ent}$ is an oscillatory function of time. The
$R(t,\tau)$ has also been calculated using Eq.(\ref{R_exp}). In the separable
case, the result does not depend on $t$ and therefore
\begin{eqnarray}
\Gamma_{sep}(\tau)= R_{sep}(t,\tau)&=& 1 + \left(2-2q^2\right)\exp\left(-q^2\right)\nonumber\\
& & + \frac{1}{2}\left[1-2q^2(s_1^2 + s_2^2)\right]\exp\left[iq^2(d_1 +
d_2)\right] \exp\left[-2q^2(s_1^2 + s_2^2)\right]
\nonumber\\
& & + \frac{1}{2}\left[1-2q^2(c_1^2 + c_2^2)\right]\exp\left[-iq^2(d_1 +
d_2)\right] \exp\left[-2q^2(c_1^2 + c_2^2)\right].
\end{eqnarray}
where $d_j = \sin(\omega_j\tau)$, $s_j = \sin(\omega_j\tau/2)$, $c_j =
\cos(\omega_j\tau/2)$ and $j=1,2$. This is a periodic function of $\tau$ only
if the ratio of $\omega_1/\omega_2$ is rational. Indeed, it can easily be
verified that if $\omega_1/\omega_2=P/Q$ where $P$ and $Q$ are coprime
integers, then the period is $(2\pi P)/\omega _1=(2\pi Q)/\omega_2$. The
$\Gamma_{sep}(\tau)$ is a quasi-periodic function of $\tau$, if the ratio of
$\omega_1/\omega_2$ is irrational. In Fig. 7 we present the absolute value of
$\Gamma_{sep}(\tau)$ as a function of $\omega_2 \tau$ for the case $\omega_1 =
1.2\times10^{-4}, \omega_2 = 10^{-4}$.

\begin{figure} [h]
\centering \scalebox{0.6}
{\includegraphics{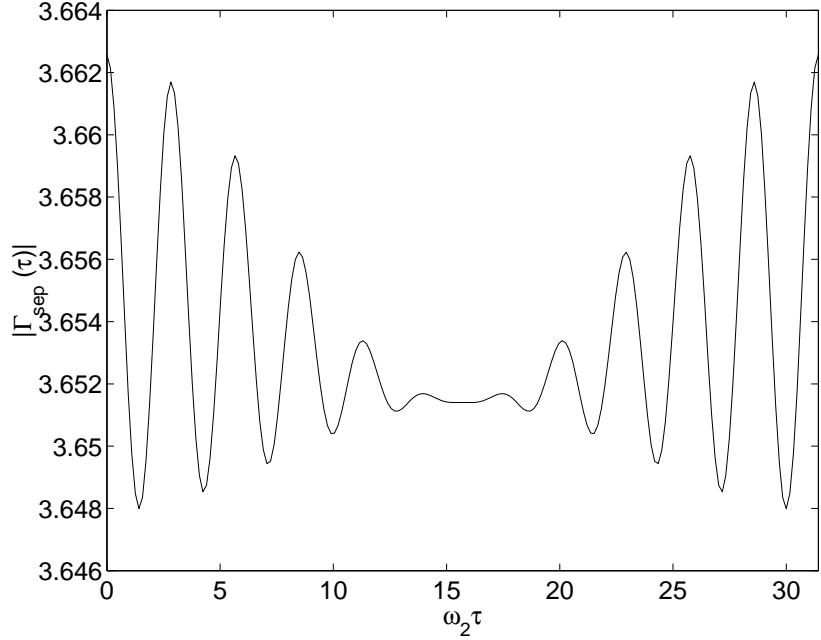}}\caption{$|\Gamma_{sep}(\tau)|$ as a function of
$\omega_2 \tau$ for the case of Eq.(\ref{sep}). We use units where
$\hbar=k_B=c=1$. }
\end{figure}

In the entangled (non-separable) microwave case
\begin{eqnarray}
R_{ent}(t,\tau) &=& R_{sep}(t,\tau) - q^2\exp\left(-q^2\right)\cos[(\omega_1-\omega_2)t]
-q^2\exp\left(-q^2\right)\cos[(\omega_1-\omega_2)(t+\tau)] \nonumber\\
& & -2q^2 s_1 s_2\exp\left[iq^2(d_1 + d_2)\right]\exp\left[-2q^2(s_1^2 + s_2^2)\right]
\cos\left[(\omega_1-\omega_2)\left(t+\frac{\tau}{2}\right)\right] \nonumber\\
& & -2q^2 c_1 c_2\exp\left[-iq^2(d_1 + d_2)\right]\exp\left[-2q^2(c_1^2 + c_2^2)\right]\cos\left[(\omega_1-\omega_2)\left(t+\frac{\tau}{2}\right)\right]
\end{eqnarray}
With regard to the periodicity of $R_{ent}(t,\tau)$ as a function of $\tau$,
similar comments can be made as for the $\Gamma_{sep}(\tau)$. We note that
$R_{sep}(t,\tau)$ is independent of $t$ while $R_{ent}(t,\tau)$ is equal to
$R_{sep}(t,\tau)$ plus an extra term which is a periodic function of $t$ with
period $(2\pi)/(\omega_1 - \omega _2)$. Therefore, integration with respect to
$t$ leads to the result that $\Gamma_{ent}(\tau) = \Gamma_{sep}(\tau)$.

\section{Discussion}
There has been a lot of work in the last few years on the interaction of
mesoscopic devices with microwaves (e.g. [6]). In this paper we have
considered non-classical microwaves which are carefully prepared in a
particular quantum state and where the quantum noise is carefully controlled.
We have studied how quantum phenomena in the microwaves, affect quantum
phenomena in the interfering electrons.

We have quantified the effect of the quantum noise on electron interference.
More specifically we have calculated both the average intensity and the
spectral density of the interference electrons for several types of
non-classical microwaves (Figs 2-5). A comparison of the results with the case
of classical microwaves demonstrates clearly the influence of the quantum
noise on the interference. The non-zero value of $\Im [\gamma(\tau)]$ in Fig 4
is a purely quantum mechanical result due to the non-commutativity of the
quantum mechanical operators $\hat I(t)$ and $\hat I(t+\tau)$. This quantity
is zero in the classical case.

We have also considered two-mode microwaves where we have shown that we get
different results for rational and irrational values of the ratio
$\omega_1/\omega_2$. We have interpreted these results in terms of emission
and absorption of photons by the non-linear device of the interfering
electrons. We have also considered both separable and entangled microwaves and
quantified their effect on the interference (Figs 6-7). The different results
in these two cases demonstrate how the deep quantum phenomenon of microwave
entanglement affects electron interference.

\section{Appendix}
The terms entering the squeezed states result in Eq.(\ref{G_sq}) are
\begin{eqnarray}
Y_1 &=& \frac{q^2}{2}[\cosh(r)-\sinh(r)\cos(2\omega t +\vartheta)] \nonumber\\
X_1 &=& 2q|B|\left[\cosh\left(\frac{r}{2}\right)\cos(\omega t -\theta_B)
-\sinh\left(\frac{r}{2}\right)\cos(\omega t +\theta_A +\vartheta)\right]\nonumber\\
Y_2 &=&\frac{q^2}{2}[\cosh(r)-\sinh(r)\cos(2\omega t + 2\omega \tau +\vartheta)] \nonumber\\
X_2 &=&2q|B|\left[\cosh\left(\frac{r}{2}\right)\cos(\omega t +\omega \tau -\theta_B)
-\sinh\left(\frac{r}{2}\right)\cos(\omega t +\omega \tau +\theta_B +\vartheta)\right]\nonumber\\
Y_3 &=&2q^2\cos^2\left(\frac{\omega \tau}{2}\right) [\cosh(r)-\sinh(r)\cos(2\omega t+\omega \tau +\vartheta)] \nonumber\\
X_3 &=&4q|B|\cos\left(\frac{\omega \tau}{2}\right)\left[\cosh\left(\frac{r}{2}\right)
\cos\left(\omega t+ \frac{\omega \tau}{2} -\theta_B\right)
-\sinh\left(\frac{r}{2}\right)\cos\left(\omega t +\frac{\omega \tau}{2} +\theta_B +\vartheta\right)\right]\nonumber\\
Y_4 &=& 2q^2\sin^2\left(\frac{\omega \tau}{2}\right) [\cosh(r)+\sinh(r)\cos(2\omega t+\omega \tau + \vartheta)] \nonumber\\
X_4 &=&4q|B|\sin\left(\frac{\omega \tau}{2}\right)\left[\cosh\left(\frac{r}{2}\right)
\sin\left(\omega t + \frac{\omega \tau}{2} -\theta_B\right)
-\sinh\left(\frac{r}{2}\right)\sin\left(\omega t +\frac{\omega \tau}{2} +\theta_B +\vartheta\right)\right]\nonumber .
\end{eqnarray}

\newpage
\section{References}
\begin{enumerate}

\item   Y. Aharonov and D. Bohm, Phys. Rev. {\bf 115}, 485 (1959).\\
        A. Tonomura \emph{et. al.}, Phys. Rev. Lett {\bf 56}, 792 (1986).\\
        M. Peshkin and A. Tonomura, \emph{The Aharonov-Bohm effect}, Lecture notes in Physics, Vol. 340 (Springer, Berlin,
        1989).

\item   S. Washburn and R.A. Webb, Adv. Phys. {\bf 35}, 375 (1986).\\
        A. G.Aronov and Y.V. Sharvin, Rev. Mod. Phys. {\bf 59}, 755 (1987).\\
        M. Pepper, Proc. Royal Soc. London A  {\bf 420}, 1 (1988).

\item   G. Badurek, H. Rauch, and J. Summhammer, Phys. Rev. Lett. {\bf 51}, 1015 (1983).\\
    J. Summhammer, Phys. Rev. A {\bf 47}, 556 (1993).\\
    J. Summhammer \emph{et al.}, Phys. Rev. Lett. {\bf 75}, 3206 (1995).

\item   A. Yacoby, M. Heiblum, D. Mahalu and H. Shtrikman, Phys. Rev. Lett. {\bf 74}, 4047 (1995).\\
        E. Buks \emph{et. al.}, Nature {\bf 391}, 871 (1998).   \\
        G. Hackenbroich, Phys. Rep. {\bf 343}, 464 (2001).

\item   M.P. Silverman, Phys. Lett. A {\bf 118}, 155 (1986); Nuovo Cimento B {\bf 97}, 200 (1987).\\
        M. Buttiker, Phys. Rev. B {\bf 46}, 12485 (1992).

\item   M. Buttiker, J. Low Temp. Phys. {\bf 118}, 519 (2000).\\
    R. Deblock \emph{et. al.}, Phys. Rev. B {\bf 65}, 075301 (2002).

\item   A. Vourdas, Phys. Rev. B {\bf 54}, 13175 (1996). \\
        A. Vourdas and B.C. Sanders, Europhys. Lett. {\bf 43}, 659 (1998). \\
    A. Vourdas, Phys. Rev. A {\bf 64}, 053814 (2001).\\
    P. Cedraschi, V.V. Ponomarenko, and M. Buttiker, Phys. Rev. Lett. {\bf 84}, 346
    (2000).

\item   R.F. Werner, Phys. Rev. A {\bf 40}, 4277 (1989).\\
        A. Peres, Phys. Rev. Lett. {\bf 77}, 1413 (1996).\\
        R. Horodecki and M. Horodecki, Phys. Rev. A {\bf 54}, 1838 (1996).\\
        V. Vedral \emph{et. al.}, Phys. Rev. Lett. {\bf 78}, 2275 (1997).

\item   S. Chountasis and A. Vourdas, Phys. Rev. A {\bf 58}, 848 (1998).

\item   A. Vourdas, Phys. Rev. B {\bf 49}, 12040 (1994).

\end{enumerate}

\end{document}